\documentclass[preprint,12pt]{elsarticle}

\usepackage{amssymb}

\usepackage{setspace}

\journal{Nuclear Physics A}

\begin{document}

\begin{frontmatter}



\title{Measurement of charged particle $R_{\rm AA}$ at high $p_{\rm T}$ in PbPb collisions at $\sqrt{s_{_{\rm NN}}}=2.76$ TeV with CMS}


\author{Kriszti\'an Krajcz\'ar on behalf of the CMS Collaboration}

\address{Massachusetts Institute of Technology, Cambridge, MA 02139, USA}

\begin{abstract}
Charged particle transverse momentum ($p_{\rm T}$) spectra have been measured by CMS for pp and 
PbPb collisions at the same $\sqrt{s_{_{\rm NN}}}=2.76$~TeV collision energy per nucleon pairs. 
Calorimeter-based jet triggers are employed to enhance the statistical reach of the high-$p_{\rm 
T}$ measurements. The nuclear modification factor ($R_{\rm AA}$) is obtained in bins of collision 
centrality for the PbPb data sample dividing by the measured pp reference spectrum. In the range 
$p_{\rm T} = 5-10$~GeV/c, the charged particle yield in the most central PbPb collisions is 
suppressed by up to a factor of 7. At higher $p_{\rm T}$, this suppression is significantly 
reduced, approaching a factor of 2 for particles with $p_{\rm T} = 40 - 100$~GeV/c.

\end{abstract}


\end{frontmatter}


\section{Introduction}

The inclusive charged particle $p_{\rm T}$ spectrum in nucleus-nucleus 
(AA) collisions is an important tool for studying high-$p_{\rm T}$ 
particle suppression in the dense QCD medium produced in high-energy AA 
collisions~\cite{DdE:JetQuenching}. The suppression (or enhancement) of 
high-$p_{\rm T}$ particles can be quantified by the ratio of charged 
particle $p_{\rm T}$ spectra in AA collisions to those in pp collisions 
scaled by the number of binary nucleon-nucleon collisions ($N_{\rm 
coll}$), known as the nuclear modification factor $R_{\rm 
AA}$~\cite{DdE:JetQuenching}:

\begin{equation}
R_{\rm AA}(p_{\rm T}) = \frac{1}{T_{\rm AA}}\frac{d^{2}N^{\rm AA}/dp_{\rm T}d\eta}{d^{2}\sigma^{NN}/dp_{\rm T}d\eta},
\label{eqn:def_raa}
\end{equation}

\noindent
where $T_{\rm AA}$ = $\langle N_{\rm coll} \rangle /\sigma^{\rm NN}_{\rm inel}$ can be calculated from a Glauber model accounting for the nuclear collision geometry \cite{Glauber}.


\section{Data samples and analysis procedure}

The measurement presented here is based on $\sqrt{s_{_{NN}}}=2.76$~TeV 
PbPb data samples corresponding to an integrated luminosities of 7~$\mathrm{\mu 
b}^{-1}$ and 150~$\mathrm{\mu b}^{-1}$, collected by the CMS experiment in 2010 
and 2011, respectively. The pp reference spectrum measured at the same 
nucleon-nucleon collision energy corresponds to an integrated luminosity 
of 230~$\mathrm{\mu b}^{-1}$.



A detailed description of the CMS detector can be found in Ref.~\cite{JINST}. The central 
feature of the CMS apparatus is a superconducting solenoid, providing a magnetic field of 
3.8~T. Immersed in the magnetic field are the silicon pixel and strip tracker, which are 
designed to provide a transverse momentum resolution of about 0.7 (2.0)\% for 1 (100) GeV/c 
charged particles at normal incidence, the lead-tungstate electromagnetic calorimeter, the 
brass/scintillator hadron calorimeter, and the gas ionization muon detectors.


In this analysis the coincidence signals of the beam scintillator counters ($3.23<|\eta|<4.65$) 
or the hadron forward calorimeters (HF; $2.9<|\eta|<5.2$) were used for triggering on minimum 
bias events. In order to extend the statistical reach of the $p_{\rm T}$ spectra, single-jet 
triggers with calibrated transverse energy thresholds were applied. The collision event 
centrality, specified as a fraction of the total inelastic cross section, is determined from 
the event-by-event total energy deposition in the HF calorimeters.

\section{Results}

\begin{figure*}
    \centering
    \includegraphics[width=0.49\textwidth]{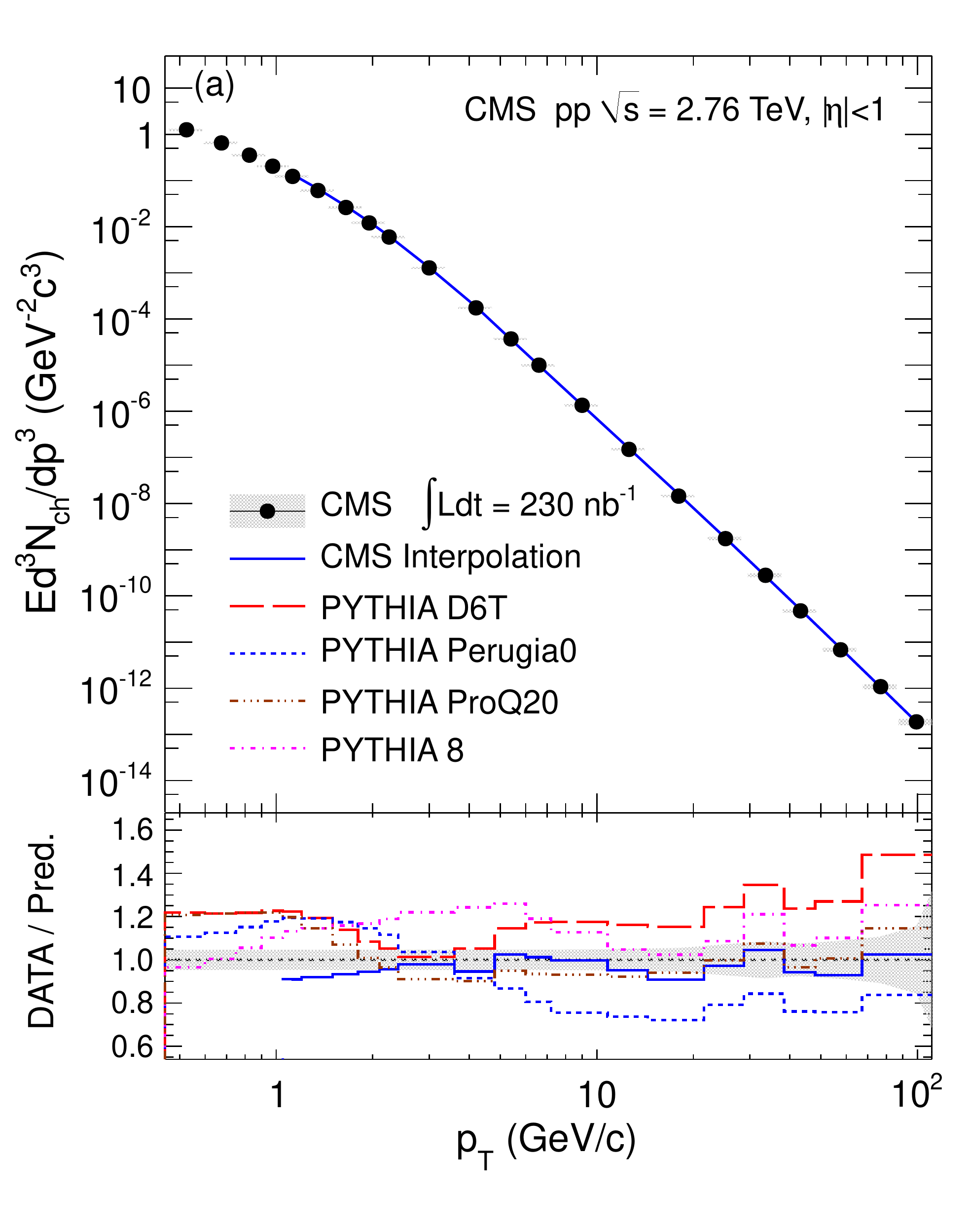}
        \includegraphics[width=0.49\textwidth]{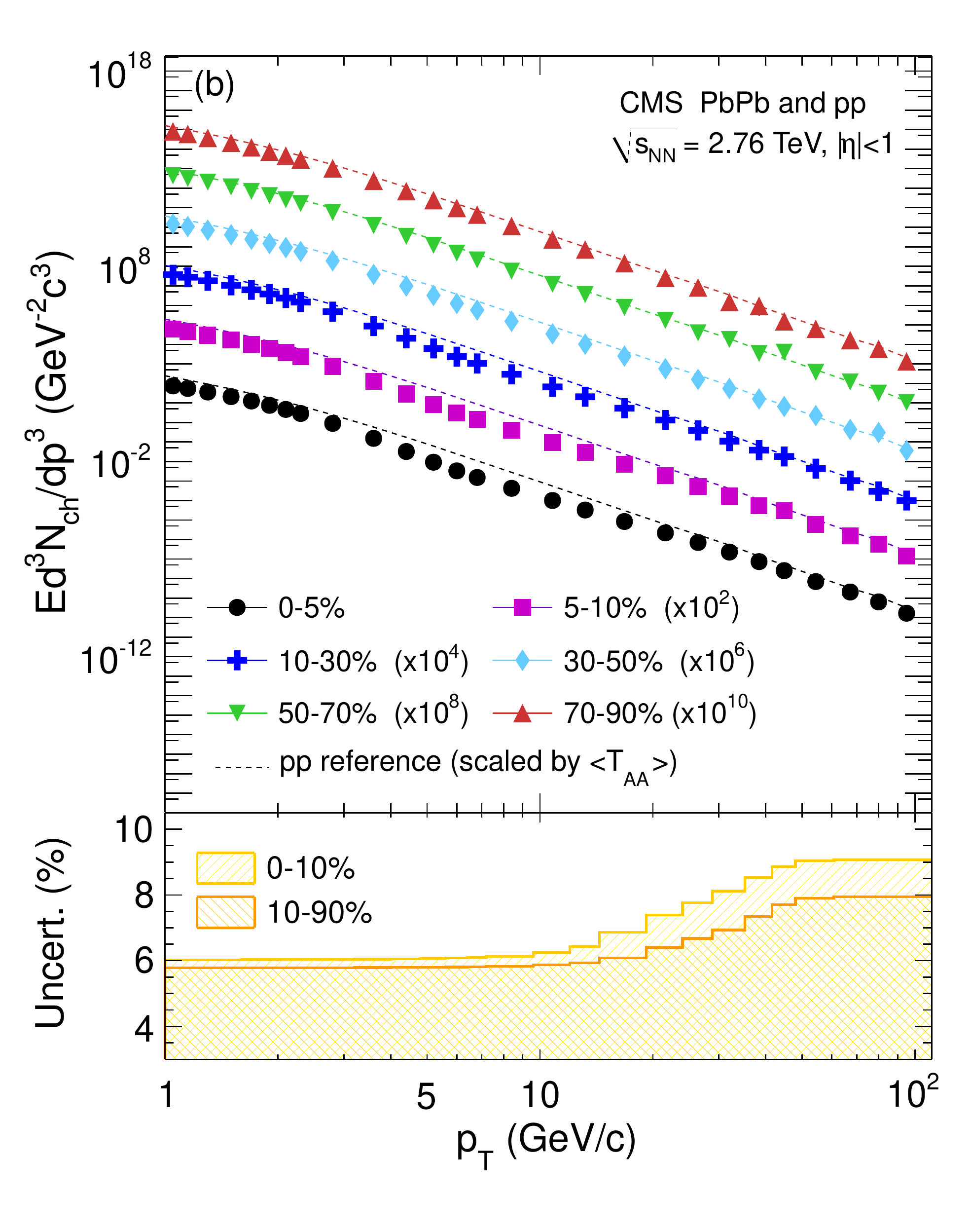}   
      \caption{\label{fig:spectra}
      (a) Upper panel: Invariant charged particle differential yield for $|\eta|<1.0$ in pp collisions at $\sqrt{s}=2.76$~TeV compared to \textsc{pythia} predictions and to the interpolated CMS spectrum~\cite{ppreferenceCMS}.
      Lower panel: the ratio of the measured spectrum to the predictions. The grey band corresponds to the statistical and systematic uncertainties of the measurement added in quadrature.
      (b) Upper panel: Invariant charged particle differential yield in PbPb collisions at 2.76~TeV in bins of
      collision centrality (symbols),
      compared to the normalized pp reference spectra (dashed lines).
      Lower panel: The relative systematic uncertainties of the PbPb differential yields
      for the 0--10\% and 10--90\% centrality intervals.
      }
\end{figure*}

The inclusive charged particle invariant differential yield averaged 
over the pseudorapidity $|\eta|<1$ in pp collisions is shown in 
Fig.~\ref{fig:spectra}(a)~\cite{CMS_raa}. Also shown are the ratios of 
the data to various generator-level predictions from the \textsc{pythia} 
MC~\cite{pythia}. The PbPb spectrum is shown in 
Fig.~\ref{fig:spectra}(b)~\cite{CMS_raa} for six centrality bins 
compared to the measured pp reference spectrum scaled by $T_{\rm AA}$. 
By comparing the PbPb measurements to the dashed lines representing the 
scaled pp reference spectrum, it is clear that the charged particle 
spectrum is strongly suppressed in central PbPb events compared to pp, 
with the most pronounced suppression at around 5--10 GeV/c.

\begin{figure*}[htb]
   \begin{center}
        \includegraphics[width=0.98\textwidth]{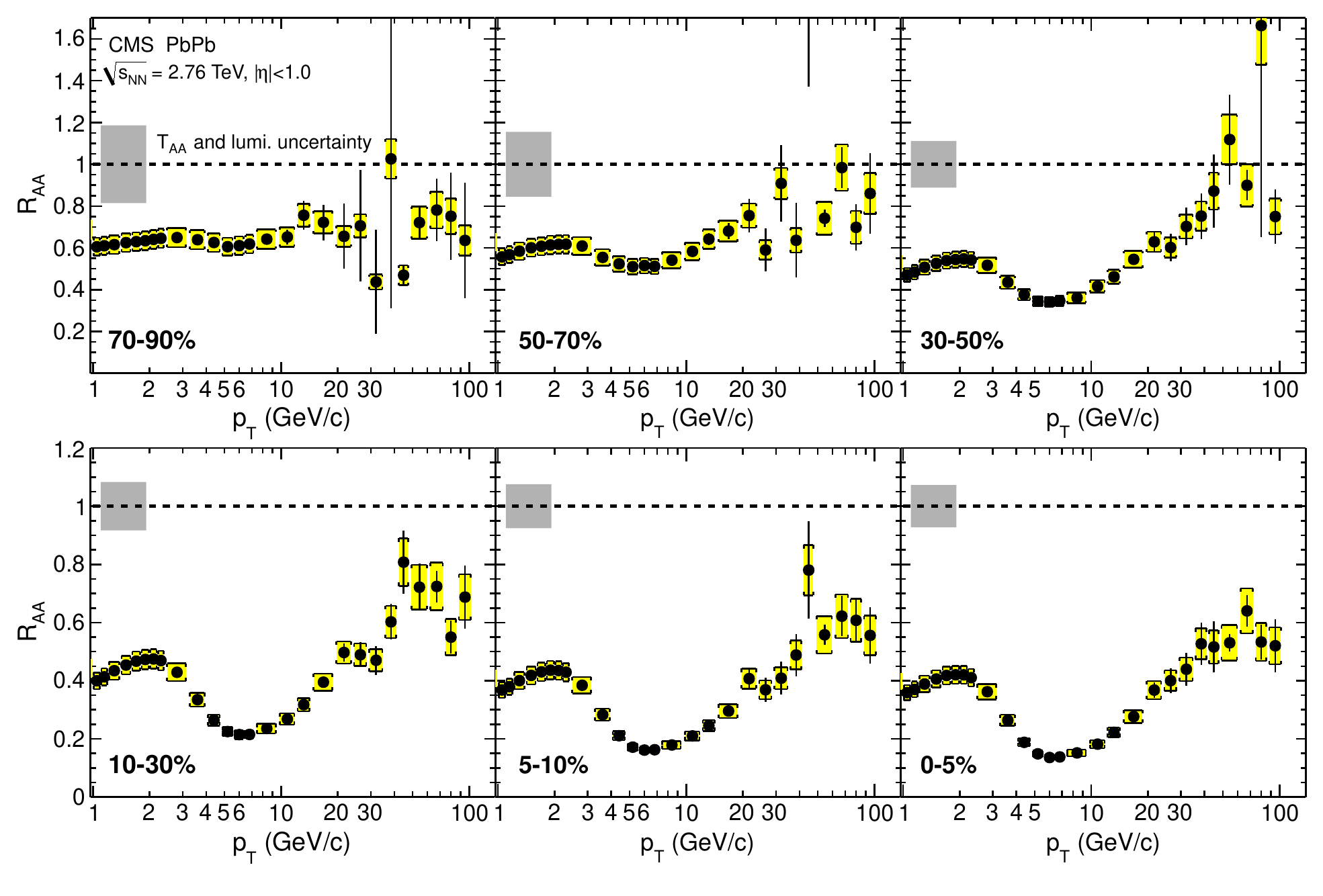}
        \caption{$R_{\rm AA}$ (black dots) as a function of $p_{\rm T}$ for six PbPb centrality bins.
        The error bars represent the statistical uncertainties and the yellow boxes represent the $p_{\rm T}$-dependent systematic uncertainties.
        An additional systematic uncertainty corresponding to the normalization factor $T_{\rm AA}$ and the pp integrated luminosity, common to all points,
        is shown as the shaded band around unity.
        }
     \label{fig:raa}
   \end{center}
\end{figure*}

The computed nuclear modification factor $R_{\rm AA}$ is shown in 
Fig.~\ref{fig:raa}~\cite{CMS_raa}. The yellow boxes around the points 
show the systematic uncertainties, including those from the pp reference 
spectrum. An additional systematic uncertainty from the $T_{\rm AA}$ 
normalization, common to all points, is displayed as the shaded band 
around unity. In case of the peripheral 70--90\% centrality bin, a moderate 
suppression of about a factor of 2 is observed at low $p_{\rm T}$, with 
$R_{\rm AA}$ rising slightly with increasing transverse momentum. The 
suppression becomes more pronounced with increasing collision 
centrality. In the 0--5\% most central centrality bin, $R_{\rm AA}$ 
reaches a minimum value of about 0.13 at $p_{\rm T}$ = 6--7~GeV/c, 
corresponding to a suppression factor of 7. At higher $p_{\rm T}$, the 
value of $R_{\rm AA}$ rises approaching roughly a suppression factor of 
2 between 40 and 100~GeV/c.

\section{Acknowledgements}
The author wishes to thank the Hungarian Scientific Research Fund
(K 81614 and NK 81447) and the Swiss National Science Foundation (128079) for their
support.




\end{document}